# Flow approaches to community detection in complex network systems

Olexandr Polishchuk

Laboratory of Modeling and Optimization of Complex Systems
Pidstryhach Institute for Applied Problems of Mechanics and Mathematics, National Academy of Sciences of Ukraine,
Lviv, Ukraine
od_polishchuk@ukr.net

**Abstract** – *The paper investigates the problem of finding communities in complex network systems, the detection of which allows a better understanding of the laws of their functioning. To solve this problem, two approaches are proposed based on the use of flows characteristics of complex network. The first of these approaches consists in calculating the parameters of influence of separate subsystems of the network system, distinguished by the principles of ordering or subordination, and the second, in using the concept of its flow core. Based on the proposed approaches, reliable criteria for finding communities have been formulated and efficient algorithms for their detection in complex network systems have been developed. It is shown that the proposed approaches make it possible to single out communities in cases in which the existing numerical and visual methods turn out to be disabled.*

## INTRODUCTION

One of the important problems that is studied in the theory of complex networks is the search for groups of interconnected nodes (clicks, clusters, communities). Identification of such groups contributes to a better understanding of the principles of organization of complex networks (CN) and the operation processes of relevant systems. In real network systems (NS), the most widespread groups are the so-called communities – subnets, the connections between the nodes of which are more numerous and stronger than between them and other nodes of the CN [1]. Examples of communities in human society are public organizations, political parties, religious denominations, national diasporas, etc., which often play a significant role in the life of their states. Many communities exist in social networks, biological and physical systems [2-4], etc.

Among the first methods for communities detection in complex network are the smallest cut, hierarchical clustering and click-based methods [5]. Algorithms based on the modularity estimation (Newman-Girvan, Blondel, Radicchi [6-8]), the spectral properties of the graph (Donetti-Munoz [9]), the estimation of network entropy (structural and dynamic methods of Rosvall-Bergstrom) and others are now widely used [10]. The main disadvantage of above mentioned algorithms for identifying communities in CN, along with computational complexity and resource consumption [11], is the lack of reliable theoretically sound criterion that defined by any of them a group of nodes actually forms a community [5, 12]. The "unreliability" of above algorithms has made popular the methods of visual search for communities [13, 14], especially in large networks. These methods are based on visual identification of CN's components, in which the density of connections is definitely higher than the density of connections in the surrounding parts of network. Obviously, the results of such search are quite subjective. The large number of existing methods for communities detection confirms the great interest to this issue and its importance.

## 1. INTEGRAL FLOW ADJACENCY MATRIX AND FLOW CORE OF NETWORK SYSTEM

Complex networks are usually described as graphs in the form $G = (V, E)$, where $V$ is the set of network nodes and $E$ is the set of connections between them. The mathematical model of CN structure is a binary adjacency matrix $\mathbf{A} = \{a_{ij}\}_{i,j=1}^{N}$, where $N$ is the number of network nodes. The values $a_{ij}$ of matrix $\mathbf{A}$ are equal to 1 if there is connection between nodes $n_i$ and $n_j$, and equal to 0 if there is no such connection. Determine the integral flow adjacency matrix $\mathbf{V}(t)$ of the volumes of flows that have passed through the network edges for the period $[t-T, t]$ up to the current time $t$

$$\mathbf{V}(t) = \{V_{ij}(t)\}_{i,j=1}^{N}, \quad V_{ij}(t) = \frac{\tilde{V}_{ij}(t)}{\max_{m,l=1,N}\{\tilde{V}_{ml}(t)\}}, \quad \tilde{V}_{ij}(t) = \int_{t-T}^{t} v_{ij}(\tau) d\tau,$$

where $v_{ij}(t)$ is the volume of flow that is on the network edge $(n_i, n_j)$ at the time $t \geq T > 0$, $i, j = \overline{1, N}$.

Matrix $\mathbf{V}(t)$, the structure of which is identical to the structure of matrix $\mathbf{A}$, is based on empirical data about the movement of flows through the network and gives a sufficiently clear quantitative view about NS operation, allows us to analyze features and predict the behavior of this process, and evaluate its effectiveness and prevent existing or potential threats [15].

Introduce the concept of flow $\lambda$-core of the network system [16], as the largest subnet of source network, for which all elements of the integral flow adjacency matrix $\mathbf{V}(t)$ have values

$$V_{ij}(t) \geq \lambda, \quad i, j = \overline{1, N}, \quad t \geq T, \quad \lambda \in [0, 1].$$

Among other things, the flow $\lambda$-core of the NS allows us to determine in its structure the most important from a functional point of view components [16].

## 2. COMMUNITIES DETECTION BASED ON SYSTEM HIERARCHIES

In real systems, the first "candidates" in the communities are the subsystems of different hierarchical levels, built on the principles of ordering or subordination [17, 18]. Let us the source network system $S$ is divided into $M$ subsystems

$$S_m \subset S = \bigcup_{m=1}^{M} S_m,$$

the sets of nodes $H_{S_m} = \{n_i^m\}_{i=1}^{N_m}$ of which do not intersect, $m = \overline{1, M}$. Denote by $G_{S_m}^{out}$ the set of all nodes-generators of flows, which are included in the set $H_{S_m}$. Determine by means of parameter

$$\xi_{S_m}^{out}(t) = \sum_{i \in G_{S_m}^{out}} \xi_i^{out}(t) / s(\mathbf{V}(t))$$

the strength of influence of subsystem $S_m$ on NS at a whole. Here $\xi_i^{out}(t)$ is a volume of output flows generated in the node $n_i$ from the set $G_{S_m}^{out}$ and

$$s(\mathbf{V}(t)) = \sum_{i=1}^{N} \sum_{j=1}^{N} V_{ij}(t)$$

is the total volume of flows that have passed through the network per period $[t–T, t]$. Let us

$$R_{S_m}^{out} = \bigcup_{i \in G_{S_m}^{out}} R_{m,i}^{out}$$

is a set of numbers of nodes which are the final receivers of flows generated in the nodes belonging to the set $G_{S_m}^{out}$. Divide the set $R_{S_m}^{out}$ into two subsets, namely

$$R_{S_m}^{out} = R_{S_m,int}^{out} \cup R_{S_m,ext}^{out},$$

where $R_{S_m,int}^{out}$ is the subset of nodes $R_{S_m}^{out}$ belonging to $H_{S_m}$, and $R_{S_m,ext}^{out}$ is the subset of nodes $R_{S_m}^{out}$ belonging to addition to $H_{S_m}$ in the source network. The set $R_{S_m,ext}^{out}$ will be called the domain of output influence of the subsystem $S_m$ on NS at a whole. The external and internal output strength of influence of the nodes-generators of flows belonging to the set $G_{S_m}^{out}$ on the subnets $R_{S_m,ext}^{out}$ and $R_{S_m,int}^{out}$ determine using the parameters

$$\xi(t)_{S_m,ext}^{out} = \sum_{i \in R_{S_m,ext}^{out}} \xi_i^{out}(t)/s(\mathbf{V}(t)),$$

$$\xi(t)_{S_m,int}^{out} = \sum_{i \in R_{S_m,int}^{out}} \xi_i^{out}(t)/s(\mathbf{V}(t))$$

Then the value

$$\omega_{S_m}^{out}(t) = \xi_{S_m,ext}^{out}(t) \Big/ \xi_{S_m,int}^{out}(t)$$

determines the relative strength of influence of subsystem $S_m$ on the network system as a whole. Namely, the smaller the value of parameter $\omega_{S_m}^{out}$, the smaller the strength of influence of subsystem $S_m$ on the NS, $m = \overline{1,M}$.

Denote by $R_{S_m}^{in}$ the set of all nodes – final receivers of flows, which are included in the set $H_{S_m}$. Determine by means of parameter

$$\xi_{S_m}^{in}(t) = \sum_{i \in R_{S_m}^{in}} \xi_i^{in}(t)/s(\mathbf{V}(t))$$

the strength of influence of network system on subsystem $S_m$, $m = \overline{1,M}$. Here $\xi_i^{in}(t)$ is a volume of input flows received in the node $n_i$ from the set $R_{S_m}^{in}$ per period $[t–T, t]$. Let us

$$G_{S_m}^{in} = \bigcup_{i \in R_{S_m}^{in}} G_{m,i}^{in}$$

is a set of numbers of nodes-generators from which flows are directed to nodes belonging to the set $R_{S_m}^{in}$. Divide the set $G_{S_m}^{in}$ into two subsets, namely

$$G_{S_m}^{in} = G_{S_m,int}^{in} \cup G_{S_m,ext}^{in},$$

where $G_{S_m,int}^{in}$ is the subset of nodes $G_{S_m}^{in}$ belonging to $H_{S_m}$, and $G_{S_m,ext}^{in}$ is the subset of nodes $G_{S_m}^{in}$ belonging to addition to $H_{S_m}$ in the source network. The set $G_{S_m,ext}^{in}$ will be called the domain of input influence of the network system on subsystem $S_m$. The external and internal input strength of influence of the nodes – final receivers of flows belonging to the set $R_{S_m}^{in}$ on the subnets $G_{S_m,ext}^{in}$ and $G_{S_m,int}^{in}$ determine using the parameters

$$\xi^{in}_{S_m,ext}(t) = \sum_{i \in G^{out}_{S_m,ext}} \xi^{in}_i(t)/s(\mathbf{V}(t)),$$

$$\xi^{in}_{S_m,int}(t) = \sum_{i \in G^{out}_{S_m,int}} \xi^{in}_i(t)/s(\mathbf{V}(t))$$

Then the value

$$\omega^{in}_{S_m}(t) = \xi^{in}_{S_m,ext}(t) \Big/ \xi^{in}_{S_m,int}(t)$$

determines the relative strength of influence of network system on subsystem $S_m$. Namely, the smaller the value of parameter $\omega^{in}_{S_m}$, the smaller the strength of influence of NS on subsystem $S_m$, $m = \overline{1,M}$.

The pair of parameters $(\omega^{out}_{S_m}, \omega^{in}_{S_m})$ forms an objective criterion of whether the subsystem $S_m$ forms a community in the network system. Indeed, the smaller the value of these parameters, the smaller the external interaction of the subsystem $S_m$ with the system as a whole and the greater the interactions within the subsystem, which is, in essence, the definition of community. We can also use the betweenness parameters of subsystem $S_m$ to build an objective criterion and corresponding algorithm for detection of communities-subsystems in the source network system, $m = \overline{1,M}$, [15].

### 3. COMMUNITIES AND FLOW CORES OF NETWORK SYSTEMS

Obviously, one of the most objective indicators of connection strength between two network nodes is the volume of flows that pass through the edge connected them over a period of time $[t-T, t]$, or in other words, the values of elements of integral flow adjacency matrix $\mathbf{V}(t), t \geq T$. This means that if during the construction of $\lambda$-core of the source NS (Fig. 1a – source CN, 1b – source NS with the reflected $\lambda$-core) with a consistent increase of value $\lambda$ at a certain value $\lambda = \lambda_1$ the flow $\lambda$-core is divided into unconnected components (Fig. 1c), then the largest communities in the network system are detected. Importantly, the structure and consist of the nodes and connections of these communities are clearly determined from the matrix $\mathbf{V}(t), t \geq T$. If with further growth $\lambda$ at a certain value $\lambda = \lambda_2$ detected in the previous step communities are again divided into unconnected components, we obtain sub-communities of these communities (Fig. 1d), etc. In contrast to the first approach, the use of flow cores of network system allows us not only to identify a particular subsystem as a community, but to perform a global search of all communities in the network system.

Note that none of numerical algorithms mentioned in the introduction, as well as the visual methods, makes it possible to detect the presence of communities in the image in Fig. 1a the simplest regular network. Similar examples can be given for much more complex real network structures [16].

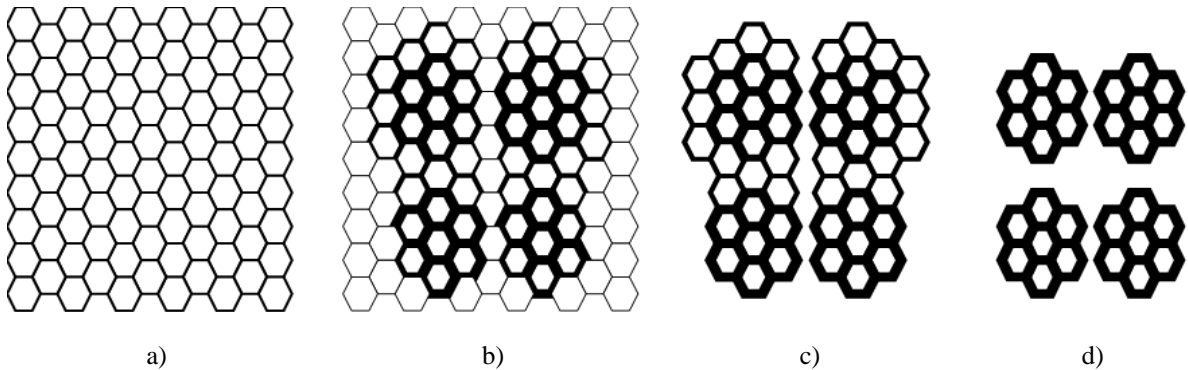

a)          b)          c)          d)

Fig. 1. Use of flow $\lambda$-cores for communities detection in complex network system

## 4. Conclusions

The paper determines the importance of problem of communities detection in complex network systems and briefly analyzes the shortcomings of known numerical and visual methods of solving this problem. Examples are shown that demonstrate the inefficiency of their use due to the lack of mathematically sound search criteria. The integral flow adjacency matrix of complex network system, parameters of influence of its separate subsystems and the concept of its flow core are defined which allowed to formulate objective criteria of communities detection in complex network and to develop effective algorithms of such detection.